\documentclass[twocolumn]{aastex62}
\usepackage{txfonts}

\shorttitle{Rocky Planetesimal Formation Aided by Organics}
\shortauthors{Homma et.al}
\newcommand{\pr}[1]{\left( #1\right)}
\newcommand{\qr}[1]{\left\{ #1\right\}}
\newcommand{\refp}[1]{(\ref{#1})}
\newcommand{\rn}{~\rm}
\newcommand{\eqref}[1]{(\ref{#1})}

\begin{document}
\title{Rocky Planetesimal Formation Aided by Organics}

\author{Kazuaki Homma}
\affiliation{Department of Earth and Planetary Sciences, 
Tokyo Institute of Technology, Meguro, Tokyo 152-8551, Japan}

\author{Satoshi Okuzumi}
\affiliation{Department of Earth and Planetary Sciences, 
Tokyo Institute of Technology, Meguro, Tokyo 152-8551, Japan}

\author{Taishi Nakamoto}
\affiliation{Department of Earth and Planetary Sciences, 
Tokyo Institute of Technology, Meguro, Tokyo 152-8551, Japan}

\author{Yuta Ueda}
\affiliation{Department of Earth and Planetary Sciences, 
Tokyo Institute of Technology, Meguro, Tokyo 152-8551, Japan}
\affiliation{Department of Earth and Planetary Science, The University of Tokyo, 7-3-1 Hongo, Tokyo 113-0033, Japan}

\correspondingauthor{Kazuaki Homma}
\email{homma.k.ac@m.titech.ac.jp}

\begin{abstract}
The poor stickiness of silicate dust grains is a major obstacle to the formation of rocky planetesimals. In this study, we examine the possibility that silicate grains with an organic mantle, which we call Organic-Mantled Grains (OMGs), form planetesimals through direct coagulation. Organic mantles are commonly found in interplanetary dust particles, and laboratory experiments show that they are softer than silicates, in particular in warm environments. This, combined with the theory of particle adhesion, implies that OMGs are stickier than bare silicate grains. Because organic mantles can survive up to 400 K, silicate grains inside the water snow line in protoplanetary disks can in principle hold such mantles. We construct a simple grain adhesion model to estimate the threshold collision velocity below which aggregates of OMGs can grow. The model shows that aggregates of 0.1 \micron-sized OMGs can overcome the fragmentation barrier in protoplanetary disks if the mantles are as thick as those in interplanetary dust particles and if the temperature is above $\sim$ 200 K. We use this {adhesion} model to simulate the global evolution of OMG aggregates in the inner part of a protoplanetary disk, demonstrating that OMG aggregates can indeed grow into planetesimals under favorable conditions. Because organic matter is unstable at excessively high temperatures, rocky planetesimal formation by the direct sticking of OMGs is expected to occur in a disk annulus corresponding to the temperature range $ \sim$ 200--400 K. The organic-rich planetesimals may grow into carbon-poor rocky planetesimals by accreting a large amount of carbon-poor chondrules. 

\end{abstract}

\keywords{dust, extinction --- planets and satellites: formation --- protoplanetary disks}
 
\section{Introduction} \label{s:intro}
The first step of planet formation is the growth of \micron-sized dust particles into km-sized planetesimals. 
Planetesimals are believed to form through the collisional growth of dust particles into larger aggregates \citep[e.g.,][]{Okuzumi2012Rapid-Coagulati,Windmark2012Planetesimal-fo,Kataoka2013Fluffy-dust-for,Arakawa2016Rocky-Planetesi} and/or the gravitational and streaming instabilities of the aggregates  \citep[e.g.,][]{Goldreich1973The-Formation-o,Johansen2007Rapid-planetesi,Youdin2011On-the-Formatio,Ida2016Formation-of-du,Drc-azkowska2017Planetesimal-fo}.
For both scenarios, 
it is crucial to understand how large the aggregates grow in protoplanetary disks.

One of the major obstacles against dust growth is the fragmentation of dust aggregates upon high-velocity collisions. 
In protoplanetary disks, macroscopic dust aggregates can collide at several tens of $\rm m~s^{-1}$ ~\citep[e.g.,][]{Johansen2014The-Multifacete,Birnstiel2016Dust-Evolution-}.
However, a number of laboratory experiments~\citep{Blum2000Experiments-on-,Guttler2010The-outcome-of-} and numerical simulations~\citep{Dominik1997The-Physics-of-,Wada2009Collisional-Gro,Wada2013Growth-efficien} have shown that aggregates made of 0.1--1~\micron-sized silicate grains are unable to stick at such a high velocity.
Therefore, it is widely believed that planetesimals do not form through 
the direct coagulation of silicate grains
(but see \citealt{Kimura2015Cohesion-of-Amo,Arakawa2016Rocky-Planetesi,Steinpilz2019Sticking-Proper} for  the possibility that silicate grains can actually be sticky).
Aggregates including water ice grains would be stickier \citep{Wada2009Collisional-Gro,Wada2013Growth-efficien,Gundlach2015The-Stickiness-}, but would only form icy planetesimals like comets. 

It is still worth asking if there are any materials that can act as a glue holding poorly sticky silicate grains together in the inner part of protoplanetary disks.
One candidate for such a material is organic matter.
One piece of evidence for the importance of organic matter comes from the observations of chondritic porous interplanetary dust particles (CP IDPs).
They are particle aggregates of likely cometary origin and represent the most primitive materials in the solar system~\citep[e.g.,][]{Ishii2008Comparison-of-C}.
Individual CP IDPs primarily consist of submicron-sized mineral and glass grains and, importantly, are bound together by organic matter mantling the individual grains \citep[e.g.,][]{Flynn1994Interplanetary-}.
Such organic-mantled grains can form when icy grain aggregates lose ice by sublimation \citep[e.g.,][]{Poch2016Sublimation-of-}.
Another piece of evidence comes from the viscoelastic measurements of interstellar and molecular-cloud organic matter analogs 
\citep{Kudo2002The-role-of-sti,Piani2017Evolution-of-Mo}. 
At temperatures above $\sim 200$~K, the analogs have low elasticity compared to silicates and also high viscosity, both of which can result in a high stickiness of the organic matter.  
In fact, \citet{Kudo2002The-role-of-sti} demonstrated that a millimeter-sized sphere of velocity $\sim 1~\rm m~s^{-1}$ can stick to a sheet of the organic matter analog at $\ga 250~\rm K$.
The organic mantle can survive up to 400 K \citep{Kouchi2002Rapid-Growth-of,Gail2017Spatial-distrib}, and therefore in principle can be present on rocky dust grains residing interior to the snow line in protoplanetary disks.

Therefore, one can anticipate a scenario in which silicate grains with organic mantles, which we call {\it Organic-Mantled Grains} (OMGs), grow into planetesimals through mutual collisions.
Although such a scenario was already proposed by \citet{Kouchi2002Rapid-Growth-of} and more recently by \citet{Piani2017Evolution-of-Mo}, detailed modeling based on the theory of protoplanetary dust evolution has never been done so far.  

In this study, we explore the possibility that OMG aggregates form planetesimals through mutual collisions in the inner part of the protoplanetary disks, in two steps.
Firstly, we model the adhesion of OMGs to study how the maximum collision velocity for sticking of two OMG aggregates depends on the grain size, temperature, and mantle thickness. 
Secondly, we simulate the global collisional evolution of OMGs in a protoplanetary disk to demonstrate that OMG aggregates can grow into planetesimals under favorable conditions.

This paper is organized as follows. 
In Section \ref{s:adhesion}, we describe our simple model for the adhesion of OMGs and explore 
the parameter dependence of the fragmentation threshold of OMG aggregates.
Section \ref{s:simulation} presents global simulations of the growth of OMG aggregates in a disk. 
In Section \ref{ss:validity}, we discuss the validity and limitations of our model and implications for terrestrial  planet formation in Section \ref{ss:imp} and  \ref{ss:how}. 
Section~\ref{s:summary} presents a summary.

\section{Modeling the Adhesion of Organic-mantled Grains}\label{s:adhesion}

\begin{figure*}[ht]
\centering
\resizebox{13cm}{!}{\includegraphics{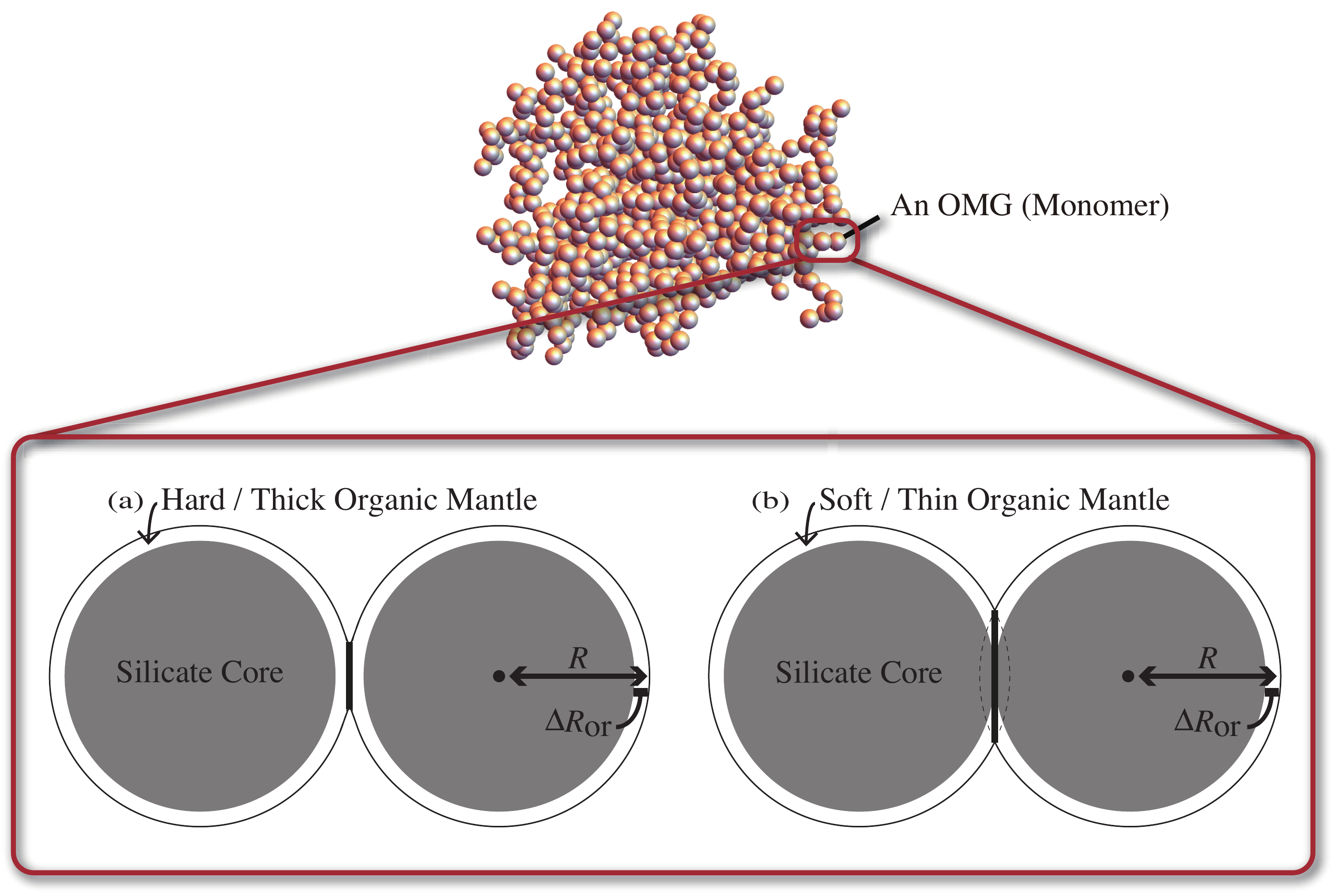}}
	    \caption{Schematic illustration of the contact of two OMGs in a dust aggregate. The gray circles and white layers represent the silicate cores and organic mantles, respectively. In the limit where the organic mantles are hard or thick (panel (a)), the silicate cores have no effect on the adhesion of the grains. In the opposite limit (panel (b)), the hard silicate cores beneath the mantle limits the size of the contact area and hence the breakup energy.} \label{f:2l}
\end{figure*}

The goal of this section is to evaluate the stickiness of aggregates made of OMGs (see Figure~\ref{f:2l} for a schematic illustration of an OMG aggregate).
In general, a collision of grain aggregates can lead to sticking, bouncing, or fragmentation depending on their collision velocity. For aggregates with porosities $\la 0.1$--0.3, sticking and fragmentation are the dominant collision outcomes \citep{Dominik1997The-Physics-of-,Blum2000Experiments-on-,Langkowski2008The-Physics-of-,Wada2011The-Rebound-Con,Meru2013Growth-and-frag}, and the collision velocity above which fragmentation dominates over sticking is called the fragmentation threshold $v_{\rm frag}$.
The fragmentation threshold is one of the key parameters that determine the fate of dust evolution and planetesimal formation \citep{Brauer2008Coagulation-fra}.

In this section, we construct a simple model that provides $v_{\rm frag}$ for OMG aggregates. 
Our model is based on the results of previous aggregates collision simulations by \citet{Wada2013Growth-efficien}, which shows that the fragmentation threshold can be estimated as
\begin{eqnarray}\label{vfrag}
 v_{\rm frag} \approx 20 \sqrt{\frac{E_{\rm break}}{m}},
\end{eqnarray}
where $m$ is the mass of individual grains (called monomers) that constitute the aggregates and $E_{\rm break}$ is the energy needed to break the contact of two monomers. 
Equation~\eqref{vfrag} implies that $v_{\rm frag}$ scales with the threshold velocity for monomer--monomer sticking, $\sim \sqrt{E_{\rm break}/m}$, which reflects the  fact that both the impact energy and total binding energy of the aggregates scale with the number of the monomers \citep[see, e.g.,][]{Dominik1997The-Physics-of-,Blum2000Experiments-on-,Wada2009Collisional-Gro}.
Thus, the breakup energy $E_{\rm break}$ is the fundamental quantity that determines the stickiness of grain aggregates.
Below we describe how we estimate 
$E_{\rm break}$ for OMGs.

In the following, we approximate each OMG as a sphere consisting of a silicate core and an organic mantle. We denote the radius of an OMG by $R$, and the thickness of the organic mantle by $\Delta R_{\rm or}$, as illustrated in Figure \ref{f:2l}. The radius of the silicate core is thus $R - \Delta R_{\rm or}$. 
The mass of an OMG is given by $m = (4\pi/3)\{\rho_{\rm sil}(R-\Delta R_{\rm or})^{3}+\rho_{\rm or}(R^3-(R-\Delta R_{\rm or})^{3})\}$, where $\rho_{\rm or}$ and $\rho_{\rm sil}$ are the material densities of the organics and silicate, respectively.
For simplicity, we assume that all monomers constituting a single aggregate are identical.

\subsection{The Breakup Energy for OMGs}\label{s:modeling}
In principle, evaluation of $E_{\rm break}$ can be done by modeling the deformation and surface attraction of particles in contact. 
An example of such a model is the Johnson--Kendall--Roberts 
(JKR) model ~\citep{Johnson301}, which has been widely used in the astronomical literature
\citep{Chokshi1993Dust-coagulatio,Dominik1997The-Physics-of-,Wada2007Numerical-Simul,Wada2009Collisional-Gro}. 
The JKR model describes the contact of two elastic particles with a positive surface energy (which represents attractive intermolecular forces) under the assumption that  
the particles have uniform internal structure. 
According to this model, the breakup energy 
for two identical spheres is given by
\begin{eqnarray}\label{JKR}
	E_{\rm break} \approx 23 
	\frac{\gamma^{5/3} R^{4/3} (1-\nu^2)^{2/3}}{Y^{2/3}},
\end{eqnarray}
where $\gamma$, $Y$, and $\nu$ are the surface energy per unit area, Young's modulus, and Poisson's ratio of the spheres, respectively.

Because the JKR model assumes uniform particles, caution is required when applying the model to OMGs, each consisting of a hard silicate core and a soft organic mantle. In general, the size of the contact area two attached particles make in  equilibrium is determined by the balance between the attractive surface forces and repulsive elastic force. 
Softer particles make a larger contact area, thus acquiring a higher breakup energy (note that $E_{\rm break}$ given by Equation~\eqref{JKR} increases with decreasing $Y$). However, a large contact area requires a large normal displacement (i.e., large deformation) of the particle surfaces. 
Now imagine the contact of two OMGs.
One can expect that the JKR model is inapplicable to OMGs if the organic mantle is so soft or thin that the normal displacement of the mantle surfaces is comparable to the mantle thickness.
In that case, the hard silicate cores should contribute significantly to the elastic repulsion between the grains, which should act to prevent the formation of a large contact area.  
Ignoring this effect would cause an overestimation of the breakup energy.

Unfortunately, there is no model that gives an exact and closed-form expression 
for $E_{\rm break}$ for particles of core--mantle structure.  
Therefore, we opt to evaluate $E_{\rm break}$ for OMGs approximately, by first considering two extreme cases and then interpolating into the intermediate regime.  

\subsubsection{The Hard/thick Mantle Regime}
We start by considering an extreme case in which the organic mantle is thick or hard enough for the silicate cores to be negligible  (see Figure \ref{f:2l}(a)). In this case, the JKR model serves as a good approximation, and hence $E_{\rm break}$ is approximately given by 
\begin{eqnarray}\label{Ethinc}
E_{\rm break} = E_{\rm or},
\end{eqnarray}
where 
\begin{eqnarray}\label{Eor}
	E_{\rm or} \approx 23 
	\frac{\gamma_{\rm or}^{5/3} R^{4/3} (1-\nu_{\rm or}^2)^{2/3}}{Y_{\rm or}^{2/3}}
\end{eqnarray}
is the breakup energy from the JKR model for the contact of uniform organics grains with radius $R$, Poisson ratios $\nu_{\rm or}$, and surface energy $\gamma_{\rm or}$ (see Equation~\eqref{JKR}).

\subsubsection{The Soft/thin Mantle Regime}

We now consider the opposite case where the organic mantles are so soft or thin that the normal displacement of the mantles are comparable to the mantle thickness (see Figure \ref{f:2l}(b)). Because the attractive surface forces try to pull the grains together, one may approximate  
the silicate cores with a soft/thin mantle 
with two attached spheres, as shown in Figure \ref{f:2l}(b). We also neglect the deformation of the mantle surface outside the contact area and approximate  the mantle surfaces with two spheres of radius $R$.
Under these approximations, the radius of the contact area is 
$\sqrt{R^2-(R-\Delta R_{\rm or})^2} = \sqrt{(2R-\Delta R_{\rm or})\Delta R_{\rm or}}$ (see Figure \ref{f:2l}(b)), and the area is 
$\pi (2R-\Delta R_{\rm or})\Delta R_{\rm or}$. 

The geometry defined above allows us to calculate 
the surface energy loss arising from the contact of the mantle surfaces. Since the surface energy per unit area is $\gamma_{\rm or}$ and the total area of the two contacting surfaces is 
$2\pi (2R-\Delta R_{\rm or})\Delta R_{\rm or}$, 
the associated energy loss is 
\begin{eqnarray} \label{Uor}
	|\Delta U_{\rm or}| = 2\pi \gamma_{\rm or} (2R-\Delta R_{\rm or})\Delta R_{\rm or}.
\end{eqnarray}
If we neglect elastically stored energy within the mantles, $|\Delta U_{\rm or}|$ gives the binding energy of the mantle contact.
The elastic energy of the soft/thin mantles is indeed negligible because mantle deformation is limited by the hard silicate cores (Young's modulus of the silicate cores is more than 10 times larger than that of the organic mantles; see Section \ref{ss:cha}).
The total binding energy $E_{\rm break}$ of the OMGs is given by the sum of $|\Delta U_{\rm or}|$ and the binding energy of the silicate cores.
In the limit of small $\Delta R_{\rm or}$, the latter contribution should approach the breakup energy of the bare silicates from the JKR model, 
\begin{eqnarray}\label{Esil}
	E_{\rm sil} \approx 23 
	\frac{\gamma_{\rm sil}^{5/3}
	(R-\Delta R_{\rm or})^{4/3} 
	(1- \nu_{\rm sil}^{2})^{2/3}
	}{Y_{\rm sil}^{2/3}}.
\end{eqnarray}
We therefore evaluate  $E_{\rm break}$  as  
\begin{eqnarray} \label{eq:Estart}
E_{\rm break}= |\Delta U_{\rm or}| +  E_{\rm sil}.
\end{eqnarray}
In fact, $E_{\rm sil}$ is negligible as long as $\Delta R_{\rm or}/R \ga 0.01$ since the silicate cores are hard and poorly sticky (see Section~\ref{ss:example}). 

\subsubsection{$E_{\rm break}$ for General Cases}
Based on the above arguments, we derive a general formula for $E_{\rm break}$ for OMGs. 
With respect to $\Delta R_{\rm or}$, We use $E_{\rm break}$ connecting equation \refp{Ethinc} and equation \refp{eq:Estart},
\begin{eqnarray} \label{modelJKR}
	E_{\rm break}= {\rm min} \qr{ E_{\rm or} , |\Delta U_{\rm or}| + E_{\rm sil} }.
\end{eqnarray}

\subsection{Material Properties}\label{ss:cha}
To evaluate $v_{\rm frag}$, one needs to assume the density, elastic constants, and surface energy of silicate and organics in addition to the grain size and mantle thickness. Here we describe our assumptions for the material properties.

\subsubsection{Silicate}
Following \citet{Chokshi1993Dust-coagulatio}, we adopt $\rho_{\rm sil} = 2.6 \rn g~cm^{-3}$,  $Y_{\rm sil} = 5.4 \times  10^{10} \rn Pa $,  $\nu_{\rm sil} = 0.17$, $ \gamma_{\rm sil} = 2.5 \times 10^{-2} ~\rm N~m^{-1}$.

\subsubsection{Organics}\label{ss:or}
We evaluate Young's modulus of organic matter using the data provided by \citet{Kudo2002The-role-of-sti}. They measured the dynamic shear modulus $G_{\rm or}$ as well as viscosity of an analog of the organic matter formed in dense molecular clouds. 
The analog is a mixture of various organic materials, including glycolic acid (see in Table 1 of \citealt{Kudo2002The-role-of-sti} for the full composition), prepared based on the analytical data of UV irradiation experiments for astrophysical ice analogs \citep{Greenberg1993Interstellar-Du,Briggs1992Comet-Halley-as}. 
We use these data because the organic mantles on protoplanetary dust particles can form in parent molecular clouds or in the disks through a similar UV irradiation process (e.g., \citet{Ciesla2012Organic-Synthes}).
The shear modulus can be translated into Young modulus if we further assume Poisson's ratio (see below).

\begin{figure} 
\centering
\resizebox{\hsize}{!}{\includegraphics{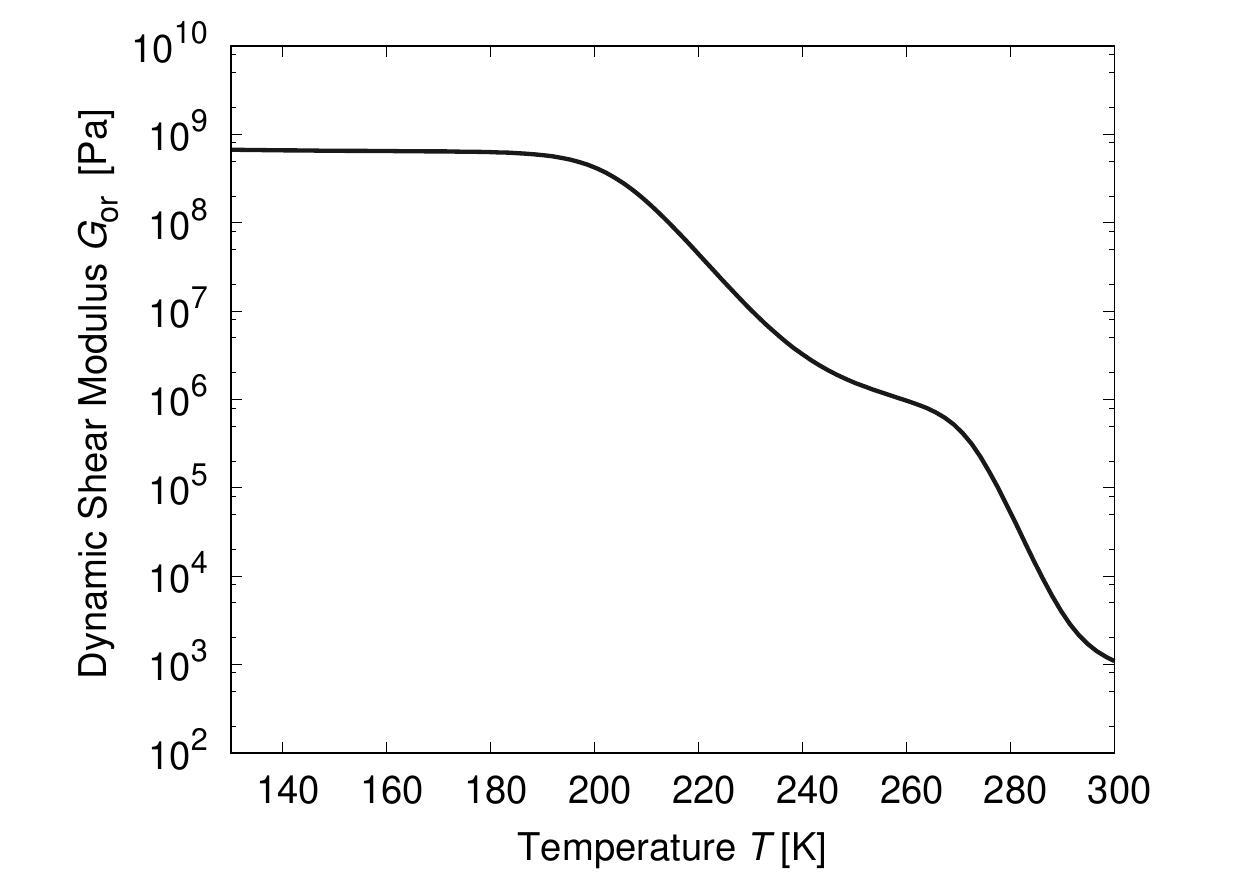}}  
\caption{Shear modulus $G_{or}$ of the organic mantles, given by Equation~\eqref{f:Gor}, as a function of temperature $T$. This is a fit to the experimental data for an analog of molecular cloud organic matter (Figure 4(a) of \citealt{Kudo2002The-role-of-sti}). \label{f:Gor}}
\end{figure}

\citet{Kudo2002The-role-of-sti} measured the shear modulus for two vibrational frequencies of 1.08 and 232 ${\rm rad~s^{-1}}$ and in the temperature range of 130--300~K, and the results are provided in Figure 4(a) of \citet{Kudo2002The-role-of-sti}.
From the data of \citet{Kudo2002The-role-of-sti} for 1.08 ${\rm rad~s^{-1}}$, we obtain an empirical fit 
\begin{eqnarray} \label{shear}
G_{\rm or} = &10&^{8.8}\frac{1 + \tanh\{(204 {\rn K}-T)/12{\rn K}\} }{2}\nonumber \\
&+&10^{9.2}\pr{\frac{1+\tanh\{(272{\rn K}-T)/7 {\rn K}\}}{2}}e^{-T/35{\rn K}} \nonumber \\
&\quad &+10^{3.1}\frac{1+\tanh\{(310 {\rn K}-T)/20 {\rn K}\}}{2}\ ~ \rm Pa,
\end{eqnarray}
which is shown in Figure \ref{f:Gor}.
Fitting to the data for 232 ${\rm rad~s^{-1}}$ yields a similar result
except in the temperature rage 200--240 K, in which the shear modulus curve 
is offset toward higher temperatures compared to the data for 1.08 ${\rm rad~s^{-1}}$ 
(see Figure 4(a) of \citet{Kudo2002The-role-of-sti}). 
Since the offset is only seen in the limited temperature range, we neglect this frequency dependence in this study. 
More discussion on this point is given in Section~\ref{ss:validity}.

The data show that $G_{\rm or}$ decreases with increasing temperature, with two plateaus $G_{\rm or} \sim 10^9~\rm Pa$ and $G_{\rm or} \sim 10^6~\rm Pa$ at $T \la 200~\rm K$ and $T \approx 240$--$270~\rm K$, respectively.
A similar temperature dependence is also commonly observed for polymers \citep[e.g.,][]{Ward2004An-Introduction}, for which the first and second plateaus in $G_{\rm or}$ are called the glassy and rubbery states, respectively \citep{Ward2004An-Introduction}. 
As we show in Section~\ref{ss:example}, the transition from the glassy to rubbery states is essential for the growth of OMGs into planetesimals. 
At $T \ga 270~\rm K$, the shear modulus further declines, attributable to the melting of the sample matter. At these temperatures, the organic matter likely behaves as a viscous fluid rather than as a solid. Viscoelastic modeling of the contact of OMGs in this temperature range will be presented in future work.

Once $G_{\rm or}$ is known, Young's modulus $Y_{\rm or}$ can be estimated using the relation $Y_{\rm or} = 2(1+\nu_{\rm or})G_{\rm or}$, where $\nu_{\rm or}$ is Poisson's ratio of the organic matter. We here assume $\nu_{\rm or} \approx 0.5$, which is the typical value for rubber.
Because Poisson's ratio generally falls within the range $0 < \nu < 0.5$ (for example, glass has $\nu \approx$ 0.2--0.3),
its uncertainty has little effect on the estimates of $Y_{\rm or}$ and $E_{\rm or}$. 
The assumed value of $\nu_{\rm or}$ gives $Y_{\rm or} = 3G_ {\rm or}$.

The surface energy $\gamma_{\rm or}$ of the organic matter of \citet{Kudo2002The-role-of-sti} is unknown, but is expected to be of the order of $10^{-2}~\rm N~m^{-1}$ as long as van der Waals interaction dominates the surface attraction force \citep[see, e.g.,][]{Chokshi1993Dust-coagulatio}. 
For this reason, we simply adopt $\gamma_{\rm or} = \gamma_{\rm sil} 
= 2.5 \times  10^{-2}~\rm N~m^{-1}$.

The material density of organics is taken to be $\rho_{\rm or} = 1.5 \rn g~cm^{-3}$. This is equal to the density of glycolic acid, one of the main constituent of the organic matter sample of \citet{Kudo2002The-role-of-sti}.
Some of the other constituents have lower densities of $\approx {1 \rn g~cm^{-3}}$, but 
this has little effect on the estimate of the grain mass (and hence $v_{\rm frag} \propto m^{-1/2}$) because the mass fraction of the organic mantle is small.
\subsection{Parameter Dependence of $v_{\rm frag}$}\label{ss:example}
Equation~\eqref{vfrag} combined with Equation~\eqref{modelJKR} gives the fragmentation threshold $v_{\rm frag}$ for OMG aggregates as a function of temperature $T$, grain size $R$, and organic mantle thickness $\Delta R_{\rm or}$. We here illustrate how $v_{\rm frag}$ depends on these parameters.

\subsubsection{Dependence on $T$ and $\Delta R_{\rm or}/R$}
\begin{figure}
\centering
\resizebox{\hsize}{!}{\includegraphics{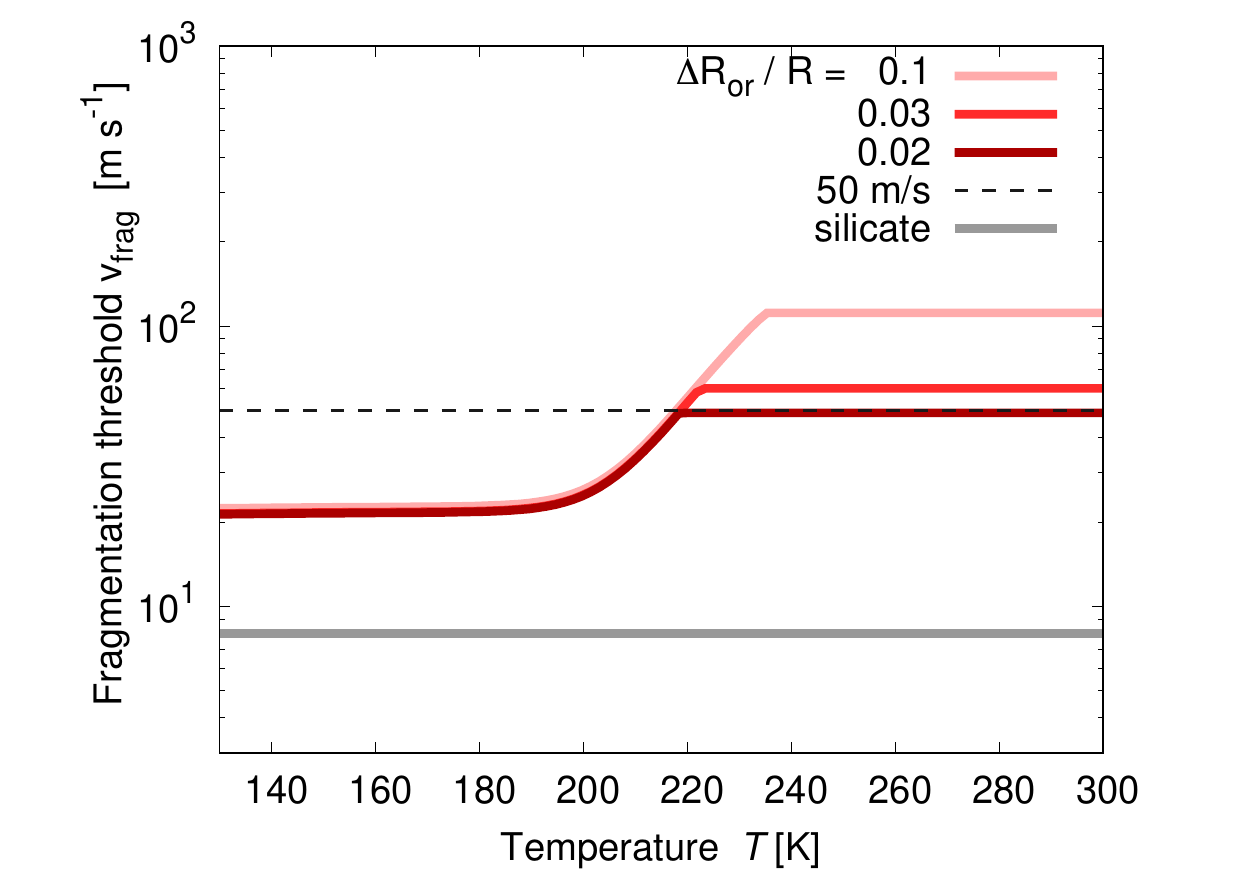}}
\caption{Fragmentation threshold $v_{\rm frag}$ for aggregates made of $0.1~\micron$-sized OMGs as a function of temperature $T$. 
The pink, red, and dark red lines are for $\Delta R_{\rm or}/R= 0.1, 0.03$ and $0.02$, respectively.
The gray line shows the fragmentation threshold for aggregates of 0.1 \micron-sized silicate grains.
The dotted line marks $50~\rm m~s^{-1}$, which is approximately equal to the maximum collision velocity in typical weakly turbulent protoplanetary disks.
}
\label{f:vfrag}
\end{figure}

\begin{figure*}[ht]
 \centering
 \resizebox{8cm}{!}{\includegraphics{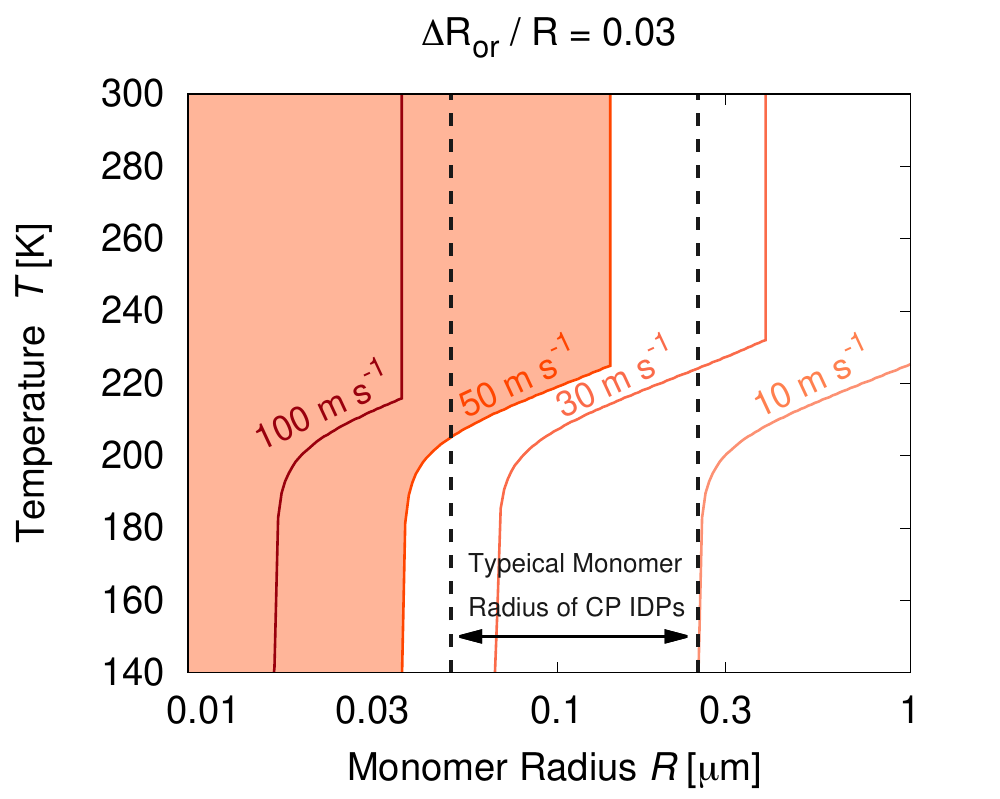}}
 \hspace{5mm}
  \resizebox{8cm}{!}{\includegraphics{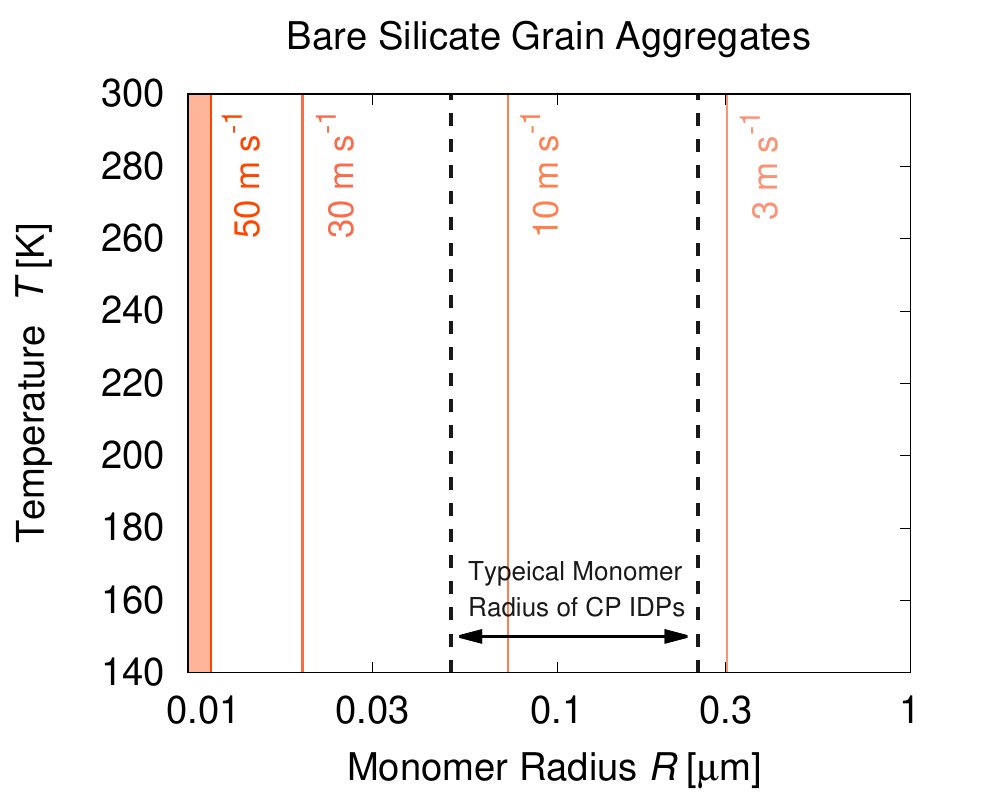}}
	    \caption{Fragmentation threshold $v_{\rm frag}$ for OMG aggregates with $\Delta R_{\rm or}/R = 0.03$ (left panel) and for bare silicate grain aggregates (right panel) as a function of monomer radius $R$ and temperature $T$. The colored area shows the parameter region in which $v_{\rm frag}$ falls below $50 \rn m~s^{-1}$, which is approximately equal to the maximum collision velocity in typical weakly turbulent protoplanetary disks. The vertical dotted lines indicate the range of the typical monomer radii of CP IDPs \citep{Rietmeijer1993Size-distributi,Rietmeijer2009A-cometary-aggr}.}\label{f:monomer1}
\end{figure*}

To begin with, we focus on the dependence on $T$ by fixing $R = 0.1~\rm \micron$ and $\Delta R_{\rm or}/R  = 0.03$.
The organic mantle thickness adopted here translates into an organic mass fraction of $\approx 5~\rm wt\%$, close to the organic content of typical IDPs  \citep[$\sim 3$--$5~\rm wt\%$;][]{Flynn2004An-assessment-o}. 
The red line in Figure \ref{f:vfrag} shows $v_{\rm frag}$ for $R = 0.1~\rm \micron$ and $\Delta R_{\rm or}/R  = 0.03$ as a function of temperature $T$.
The temperature dependence comes from that of $Y_{\rm or} \propto G_{\rm or}$ involved in the expression of $E_{\rm break}$ in the hard mantle regime (Equation~\eqref{Eor}).  
As shown in Figure~\ref{f:Gor}, the shear modulus of the organic mantle is constant at $T \la 200\rn K$ and starts to decline above this temperature range.
Accordingly, the fragmentation threshold for hard mantles, $v_{\rm frag} \propto E_{\rm or}^{1/2} \propto G_{\rm or}^{-1/3}$, increases at $T \ga 200~\rm K$. The increase in $v_{\rm frag}$ stops at $T\approx 220~\rm K$, at which the breakup energy reaches the soft-mantle limit set by the finite thickness of the organic mantle, Equation~\eqref{eq:Estart}. 
In the particular case shown here, the contribution $|\Delta U_{\rm or}|$ from the mantle interface gives the dominant contribution to $E_{\rm break}$, with the core contribution $E_{\rm sil}$ being 60 times smaller than $|\Delta U_{\rm or}|$.

The mantle thickness $\Delta R_{\rm or}$ controls the maximum fragmentation threshold attained in the soft/thin limit. The pink and dark red lines in Figure~\ref{f:vfrag} show $v_{\rm frag}$ for $\Delta R_{\rm or}/R  = 0.1$ and 0.02, respectively.
In the soft/thin limit, the fragmentation threshold scales as 
$v_{\rm frag} \propto |\Delta U_{\rm or}|^{1/2} \propto \Delta R_{\rm or}^{1/2}$, where we have assumed $|\Delta U_{\rm or}| \gg E_{\rm sil}$ 
and $\Delta R_{\rm or} \gg R$. 
For $\Delta R_{\rm or}/R =$ 0.1, 0.03, and 0.02, 
the maximum fragmentation threshold is 
$v_{\rm frag} \approx 110, 60$ and $49 \rn m ~ s^{-1}$, respectively.

The results shown here have two important implications.
Firstly, OMG aggregates are stickier than aggregates of bare silicates, even at low temperatures. 
The fragmentation threshold for bare silicate aggregates is $v_{\rm frag} = 20\sqrt{E_{\rm sil}/m}$, 
which is $\approx 8~\rm m~s^{-1}$ for $R = 0.1~\micron$
(see the gray line in Figure~\ref{f:vfrag}). 
This is smaller than the threshold for OMG aggregates in the low temperature limit, $v_{\rm frag} \approx 22~\rm m~s^{-1}$. 
This is because the organic mantles at low temperatures is still softer than the silicate cores: $Y_{\rm or}  = 2.7\times 10^9 ~\rm Pa$ versus $Y_{\rm sil} = 5.4\times 10^{10}~\rm Pa$.  
Secondly, the fragmentation threshold of OMGs can exceed $50 \rn m~s^{-1}$, which is the maximum collision velocity of dust aggregates in weakly turbulent protoplanetary disks \citep[see, e.g., Figure 1 of][]{Birnstiel2016Dust-Evolution-}.
Therefore, the OMG aggregates may grow to be planetesimals through collisions. 
For $R = 0.1~\micron$, $v_{\rm frag}$ exceeds $50~\rm m~s^{-1}$ if $\Delta R_{\rm or}/R \ga 0.03$.
\subsubsection{Dependence on $R$}
In general, the fragmentation threshold 
$v_{\rm frag} \propto \sqrt{E_{\rm break}/m}$
decreases with increasing the monomer size $R$ because $E_{\rm break}$ increases more slowly than the monomer mass $m$ $(\propto R^3)$.
To see this, we plot in the left panel of Figure~\ref{f:monomer1} the fragmentation threshold as a function of $R$
and $T$ for OMG aggregates with $\Delta R_{\rm or}/R= 0.03$.
For comparison, the fragmentation threshold for bare silicate aggregates is also shown in the right panel of Figure~\ref{f:monomer1}. 
As long as the JKR theory is applicable to the organic mantle (the hard/thick mantle limit), $v_{\rm frag}$ scales with $R$ as $R ^{-5/6}$. This scaling can be seen in the left panel of Figures~\ref{f:monomer1} at $T \la 220~\rm K$, and in the entire part of the right panel of Figure~\ref{f:monomer1} because the JKR theory applies to bare silicate grains.
At higher temperatures, at which the soft/thin mantle limit applies, the scaling becomes 
$v_{\rm frag} \propto R^{-1} \Delta R_{\rm or}^{1/2}$ 
because $E_{\rm break} \approx |\Delta U_{\rm or}| \propto R \Delta R_{\rm or}$.  

Figure \ref{f:monomer1} can be used to infer the parameter range in which OMG aggregates can grow beyond  the fragmentation barrier. Taking $50~\rm m~s^{-1}$ as a representative value for the maximum collision velocity in weakly turbulent protoplanetary disks, the fragmentation threshold exceeds the maximum collision velocity in the color shaded regions in Figure \ref{f:monomer1}. 
For comparison, the dotted lines in Figure \ref{f:monomer1} mark the sizes of grains that dominate the matrix of IDPs, $R \approx 0.05$--$0.25 ~\micron$ (diameters of 0.1--0.5 \micron; e.g., \citealt{Rietmeijer1993Size-distributi,Rietmeijer2009A-cometary-aggr}).
We find that aggregates made of OMGs of $R \approx 0.05$--$0.14 \rn \micron$ can grow beyond the fragmentation barrier as long as the temperature is above $200~\rm K$.
In contrast, bare silicate grain aggregates (the left panel of Figure~\ref{f:monomer1}) can only overcome the fragmentation barrier if the monomer radius is as small as $R \la 0.01~\micron$ \citep[see also][]{Arakawa2016Rocky-Planetesi}

\section{Global Evolution of OMG Aggregates} \label{s:simulation}

In Section~\ref{ss:example}, we showed that OMG aggregates can break through the fragmentation barrier if the temperature is high and/or the grains are sufficiently small. 
In this section, we present simulations of the size evolution of OMG aggregates in protoplanetary disks to demonstrate that the OMG aggregates indeed form planetesimals under favorable conditions.

\subsection{Model}\label{ss:simulation methods}

\subsubsection{Disk Model}
We adopt the minimum-mass solar nebula (MMSN) model~\citep{Hayashi1981Structure-of-th} around a solar-mass star as a model of protoplanetary gas disks.
In this model, the gas surface density is given by $\Sigma_{\rm g} = 1700(r/1~\rm au)^{-3/2}\rn g~ cm^{-2}$, where $r$ is the distance from the central star.
Assuming hydrostatic equilibrium and uniform temperature in the vertical direction, the gas density is given by $\rho_{\rm g} = \Sigma_{\rm g }/(\sqrt{2\pi}h_{\rm g})\exp (-z^{2}/{(2h_{\rm g}^2)})$, where $h_{\rm g} = c_{s}/\Omega$, is the scale height with $c_{s}$ and $\Omega$ being the  sound velocity and Keplerian frequency, respectively.
The gas temperature is simply set to be $T = 280(r/1\rn au)^{-1/2}~K$, which corresponds to the temperature profile in an optically thin disk around a solar-luminosity star \citep{Hayashi1981Structure-of-th}.
In reality, dusty protoplanetary disks are optically thick, and the gas temperature  at the midplane may be higher or lower than assumed here depending on the distribution of turbulence inside the disks \citep{Hirose2011Heating-and-Coo,Mori2019Temperature-Str}.
Thus, the temperature model adopted here should be taken as one reference model. 
The initial dust-to-gas mass ratio is taken to be 0.01.

In the simulations, the radial computational domain is taken to be $0.5 {\rn au} \leq r \leq 3 {\rn au}$.
The outer edge of the computational domain is equivalent to the snow line in our disk model.

\subsubsection{Dust Model}
We follow the global evolution of OMG aggregates in the modeled disk using the bin scheme described in \citet{Okuzumi2012Rapid-Coagulati}. This scheme, originally developed by \citet{Brauer2008Coagulation-fra}, evolves the full size distribution of OMG aggregates at different radial locations by taking into account the coagulation/fragmentation and radial transport of the aggregates. We refer to Sections 2.2 and 2.4 of \citet{Okuzumi2012Rapid-Coagulati} for details on the algorithm and numerical implementation.  
The dust in the disk is assumed to be initially in the form of detached OMGs of equal radius $R = 0.1~\micron$ and equal organic-mantle thickness $\Delta R_{\rm or}$.
The mantle thickness is taken to be either $\Delta R_{\rm or}/R = 0.02$ or $0.03$ to demonstrate that breaking through the fragmentation barrier requires a sufficiently thick organic mantle (see Section~\ref{ss:example}).
The OMG grains are assumed to grow into aggregates of constant internal density $\rho_{\rm int} = 0.1\rho_{\rm m}$, where $\rho_{\rm m}$ is the density of individual OMGs. For $\Delta R_{\rm or}/R = 0.03$, we have $\rho_{\rm m} = 2.5 \rn g~cm^{-3}$, which gives $\rho_{\rm int} = 0.25 \rn g~cm^{-3}$. This assumed internal density is comparable to those of CP IDPs \citep[e.g.,][]{Rietmeijer1993Size-distributi}.

 \begin{figure*}[t]
\centering
 \resizebox{8cm}{!}{\includegraphics{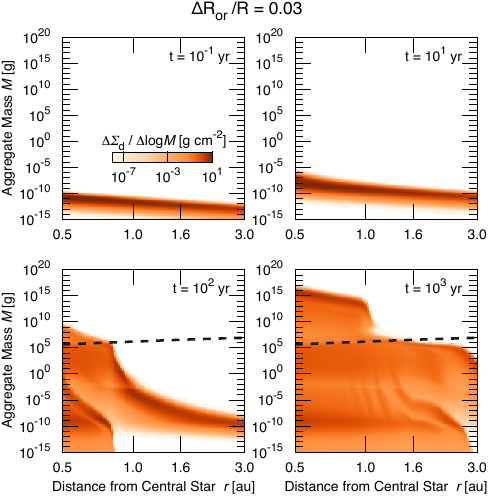}}
 \hspace{5mm}
  \resizebox{8cm}{!}{\includegraphics{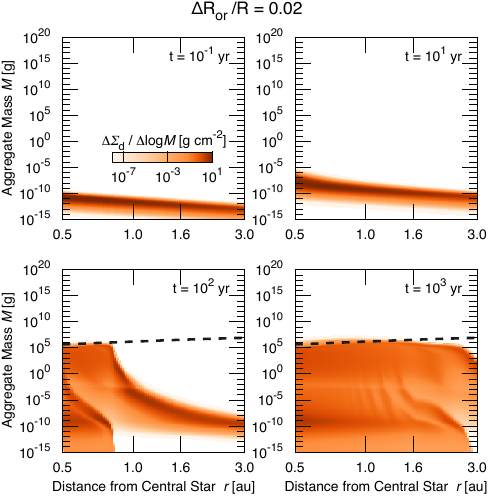}}
  \caption{Evolution of the size distribution $\Delta \Sigma_{\rm d} /\Delta \log m$ of OMG aggregates for $\Delta R_{\rm or}/R = 0.03$ and 0.02 (left and right panels, respectively). The dotted line shows the aggregate size at which the radial drift velocity takes its maximum. Near this size, the collision velocity of the aggregates is also maximized.  }
  \label{f:grow32}
\end{figure*}

Because of the sub-Keplerian motion of the gas disk, macroscopic aggregates undergo radial inward drift \citep{Adachi1976The-gas-drag-ef,Weidenschilling1977Aerodynamics-of,Whipple1972On-certain-aero}, 
which is included in our model.
The magnitude of the radial drift velocity, which depends on aggregate size, is evaluated using Equations (4)--(7) of \citet{Okuzumi2012Rapid-Coagulati}.
In our disk model, the drift speed is maximized when the aggregate mass is $\approx 4.7 \times 10^5$ g at $r = 0.5~\rm au$ and $\approx 8.9 \times 10^6$ g at $r =3~\rm au$, respectively.

The outcome of aggregate collisions is determined using a simple model adopted by \citet{Okuzumi2012Planetesimal-Fo}.
This model assumes that when two aggregates of masses $M_{l}$ and $M_{s} (<M_l)$ collide at velocity $\Delta v$, a new aggregate of mass $M = M_{l} + s(\Delta v, v_{\rm frag})M_{s}$ forms, where
$ s(\Delta v,v_{\rm frag})$ is the dimensionless sticking efficiency depending on the ratio between $\Delta v$ and the fragmentation threshold $v_{\rm frag}$ (see below).
The rest of the mass of the two collided aggregates, $M_l+M_s-M = (1-s)M_s$, goes to small fragments.
For simplicity, we neglect the size distribution of the fragments by assuming that all fragments are small as the initial OMGs (i.e., $R=0.1~\micron$).
This assumption overestimates the number of tiny aggregates, but little affects the growth of the largest aggregates because $\Delta v$ is generally controlled by the larger of two colliding aggregates.
The fragmentation threshold for OMG aggregates is determined as a function of $\Delta R_{\rm or}/R$ and $T$ using Equations~\eqref{vfrag} and \eqref{modelJKR}.

The sticking efficiency $ s(\Delta v, v_{\rm frag})$ is evaluated as
\begin{eqnarray}\label{e:s}
s(\Delta v, v_{\rm frag}) = {\rm min} \left\{ 1, -\frac{ \ln (\Delta v/v_{\rm frag})}{\ln 5} \right\}.
\end{eqnarray}
This, originally proposed by \citet{Okuzumi2012Planetesimal-Fo}, is a fit to the results of aggregate collision simulations by \citet{Wada2009Collisional-Gro}.
If the large and small aggregates collide at $\Delta v \ll v_{\rm frag} $, these aggregates stick perfectly~($s(\Delta v, v_{\rm frag})= 1$).
If aggregates collide at $\Delta v \approx v_{\rm frag} $, a part of them are released as fragments ($0< s(\Delta v, v_{\rm frag})< 1$).

The collision velocity $\Delta v$ 
includes the contributions from Brownian motion~\citep{Nakagawa1986Settling-and-gr}, radial and azimuthal drift, vertical settling~\citep{Adachi1976The-gas-drag-ef, Weidenschilling1977Aerodynamics-of} and gas turbulence
 ~\citep{Ormel2007Closed-form-exp}, which we compute using Equations (16)--(20) of \citet{Okuzumi2012Rapid-Coagulati}. 
The turbulence-driven collision velocity depends on a dimensionless parameter $\alpha$ that quantifies the strength of turbulence, and we take $\alpha = 10^{-3}$ in this study.
The aggregate size at which the collision velocity is maximized is approximately equal to the size at which the radial drift velocity is maximized. In particular, for collisions of different-sized aggregates,
the maximum collision velocity is approximately equal to the maximum speed of radial drift, which is $\approx 54~\rm m~s^{-1}$ in our disk model.

\subsection{Results}\label{ss:simulation results}
Figure \ref{f:grow32} presents our simulation results for $\Delta R_{\rm or}/R = 0.03$ and 0.02, respectively. 
The figure shows the radial distribution of aggregate surface density per decade in aggregate mass,  $\Delta \Sigma_{\rm d}/ \Delta\log{M}$, at different times $t$ after the beginning of the simulations.
The dotted line marks the aggregates mass at which the radial drift becomes fastest. This line can be regarded as the peaks of the drift and fragmentation barriers because the collision velocity takes its maximum near this size.

The results shown in Figure~\ref{f:grow32} demonstrate that OMG aggregates in the inner warm region of protoplanetary disks can form km-sized planetesimals if the organic mantle is sufficiently thick.
For $\Delta R_{\rm or}/R = 0.03$, the left panels of Figure \ref{f:grow32} show that growth beyond the fragmentation barrier occurs at $r \la 1.6 \rn au$. 
In this region, the gas temperature exceeds $220~\rm K$, so that the fragmentation threshold exceeds the maximum collision velocity $\sim 50~\rm m~s^{-1}$ as already seen in Figure~\ref{f:vfrag}. 
Within $10^3$ years, the mass of the largest aggregates reaches $10^{15}~\rm g$, which amounts to the mass of km-sized solid bodies.
At $r \ga 1.6~\rm au$, the growth of OMG aggregates stalls at the size at which the collision velocity reaches the fragmentation threshold in this region, $\approx 20~\rm m~s^{-1}$. 
In this particular simulation, the maximum aggregate mass in this region is $\sim 10^5$--$10^6$ g, corresponding to aggregate radii of $\sim 1~\rm m$.
For $\Delta R_{\rm or}/R = 0.02$, the right panels of Figure \ref{f:grow32} show that fragmentation limits the growth of OMG aggregates even in the inner warm region of the disk, again confirming our prediction. 

It is interesting to note that the OMG aggregates that have overcome the fragmentation barrier also break through the radial drift barrier, i.e., they grow faster than they fall toward the central star, at $r \la 1.0 \rn au$. As we explain below, this is a purely aerodynamic effect, not directly related to the stickiness of OMGs. In this inner disk region, the gas drag acting on rapidly drifting aggregates follows the Stokes drag law, i.e., the aggregates are larger than the mean free path of gas molecules. In this case, the aggregate size at which the fastest radial drift occurs decreases with decreasing $r$  \citep[e.g.,][see also our Figure~\ref{f:grow32}]{Birnstiel2010Gas--and-dust-e,Okuzumi2012Rapid-Coagulati,Drc-azkowska2014Rapid-planetesi}. For this reason, aggregates at $r \la 1.0 \rn au$ are able to overcome the drift barrier unless they experience mass loss due to fragmentation (see, e.g., Figure 11 of \citealt{Birnstiel2010Gas--and-dust-e}). This aerodynamic effect depends on the assumed porosity of the aggregates  \citep{Okuzumi2012Rapid-Coagulati}; for $\rho_{\rm int} = 0.01\rho_{\rm m}$, we find that the OMG aggregates overcome the radial drift barrier already at $r \approx 1.6~\rm au$.

\section{Discussion}\label{s:disussion}

\subsection{Validity and Limitations of the Adhesion Model}\label{ss:validity}
As illustrated in Section~\ref{ss:example}, one of the most important quantities that determine the stickiness of OMGs is Young's modulus of their organic mantles, in particular its dependence on temperature.  
The transition to the rubbery ($Y_{\rm or} \sim 10^6~\rm Pa$) state in warm environments is essential for aggregates of the OMGs to grow even at the maximum collision velocity in protoplanetary disks (Figures~\ref{f:vfrag} and \ref{f:grow32}). 
The question is then whether such temperature dependence is peculiar to the organic matter of \citet{Kudo2002The-role-of-sti} or can commonly be observed for organics formed by UV irradiation. 
\citet{Piani2017Evolution-of-Mo} recently measured the viscoelasticity of their own molecular cloud organic matter analogs at room temperature, showing that Young's modulus measured at a frequency of 1 Hz is $\sim 10^6~\rm Pa$, 
similar to the value for the organic matter of \citet{Kudo2002The-role-of-sti} at $T \approx 260~\rm K$. 
Therefore, we expect that the glassy-to-rubbery transition leading to the breakthrough of the fragmentation barrier occurs at $T\approx 260$--$300~\rm K$, even if the mechanical properties of real organic matter in protoplanetary disks are highly uncertain. 

One potentially important uncertainty in the elasticity of the organic mantles is its dependence on dynamical timescale. 
Our model for the shear modulus of organics (Equation~\eqref{f:Gor}) is based on the viscoelastic measurements at frequencies of 1.08 and 232 ${\rm rad~s^{-1}}$ (0.17 and 37 Hz, respectively), thus best representing the elastic properties on dynamical timescales of $10^{-2}$--$10~\rm s$.
These timescales are, however, far longer than the typical timescale of the deformation of aggregates that collide in protoplanetary disks.
For a collision velocity of $\approx 50 \rn m~s^{-1}$ and an organic-mantle thickness of $\Delta R_{\rm or}\approx 0.003 ~\micron$, the timescale on which the organic mantles inside the colliding aggregates deform is estimated to be as short as $\sim 10^{-10} ~\rm s$. 
The viscoelastic data of \citet[][their Figure 4]{Kudo2002The-role-of-sti} seem to imply that the shear modulus in the temperature range $200$--$240~\rm K$ increases slowly with increasing frequency. 
A similar trend can also be seen in the viscoelastic data for the molecular cloud organic analogs by \citet[][their Figure 7(a)]{Piani2017Evolution-of-Mo} measured at room temperature and in the frequency range $1$--250 Hz.
Therefore, using elasticity data at the low frequencies might result in an overestimation of the stickiness of protoplanetary OMGs.
However, if such frequency dependence is only limited to a narrow temperature range as expected from the data of \citet{Kudo2002The-role-of-sti}, it would not affect our main conclusion that OMG aggregates can break through the fragmentation barrier in warm regions of disks.

Finally, we note that the viscosity of the organic matter, which is not included in our grain contact model, can further enhance the stickiness of OMGs as expected by \citet{Kudo2002The-role-of-sti}. 
In principle, this effect can be evaluated by using a contact model that treats viscoelasticity \citep[e.g.,][]{Krijt2013Energy-dissipat} together with the viscosity data obtained by \citet{Kudo2002The-role-of-sti}. We leave this for future work.

\subsection{Rocky Planetesimal Formation in a Narrow Annulus? }\label{ss:imp}
We have shown that OMGs in warm regions of protoplanetary disks can grow into planetesimals. 
In fact, however, dust growth facilitated by organics is unlikely to occur in hot regions of $T \ga$ 400 K, because organic mantles are no longer stable at such temperatures.
For example,
\citet{Kouchi2002Rapid-Growth-of} report that their molecular cloud organic matter analog evaporated at approximately 400 K.
\citet{Gail2017Spatial-distrib} also point out that pyrolysis of organic materials occurs at 300--400 K.
Thus, we can only expect planetesimal formation through the direct sticking of OMGs to occur in a certain temperature range (e.g., $T \sim 200$--400 K for the organic matter analog of \citealt{Kouchi2002Rapid-Growth-of}) where the organic mantles are stable and also soft enough to bind silicates together. 
Because the disk temperature is a decreasing function of the orbital radius, the above implies that planetesimal formation aided by organics only occurs in a certain range of orbital radii, as first pointed out by   \citet{Kouchi2002Rapid-Growth-of}.

Interestingly, such a narrow planetesimal-forming zone can provide favorable conditions for the formation of the terrestrial planets in the solar system.  
\citet{Kokubo2006Formation-of-Te} and \citet{Hansen2009Formation-of-th} showed using $N$-body simulations of planetary accretion that protoplanets initially placed in an  annulus around 1 au form a planetary system whose final configuration resembles that of the inner solar system.  
However, the initial protoplanets in their simulations, which are $\sim 0.01$--$0.1M_\oplus$ in mass, are much larger 1--100 km planetesimals.
The simulations of planetesimal formation that we showed in the previous section do not include gravitational scattering and focusing between solid bodies, both of which are important for the growth and orbital evolution of planetesimal-sized bodies. Dynamical simulations bridging the gap between the planetesimals and protoplanets are needed to assess whether dust growth facilitated by organics can indeed result in the formation of planetary systems like the inner solar system.

\subsection{How Can Carbon-poor Rocky Planetesimals Form?}\label{ss:how}
Another issue in relating our planetesimal formation scenario to the formation of the solar-system terrestrial planets is the low carbon content of the Earth, and possibly of other terrestrial planets. The carbon content of the Earth's bulk mantle is estimated to be $\approx 0.01$--$0.08$~wt\%~\citep{McDonough1995The-composition,Marty2012The-origins-and,Palme20143.1---Cosmochem}.  
Some Martian meteorites contain magmatic carbon of $\sim 10^{-4}$--$10^{-2}$ wt\% \citep{Grady2004Magmatic-carbon}, possibly pointing to a carbon abundance in the Martian mantle as low as that in the Earth's mantle.
The cores of terrestrial planets could have a higher carbon abundance because carbon is highly soluble in liquid iron \citep{Wood1993Carbon-in-the-c,Wood2013Carbon-in-the-C,Chi2014Partitioning-of,Tsuno2018Core-mantle-fra}, although the carbon abundance of the Earth's core is unlikely to be much in excess of 1 wt\% \citep{Nakajima2015Carbon-depleted}.
Even if all the terrestrial carbon had been delivered in the form of organic mantles, its abundance would not have been high enough for the organic-mantled dust grains to grow beyond the fragmentation barrier, which requires an organic content of $\ga 5~\rm wt\%$ (see Section~\ref{ss:example}).
Although it is possible that rocky protoplanets lose some amount of volatiles including carbon when they grow through giant impacts  \citep[e.g.,][]{Genda2005Enhanced-atmosp}, it seems unlikely that the carbon content of the protoplanets that formed the Earth was many orders of magnitude higher than that of the present-day Earth.

However, it is possible that the carbon content of the planetesimals that formed the terrestrial planets decreased when they grew by accreting a large amount of carbon-poor solid particles.
A piece of evidence for this comes from ordinary and enstatite chondrites, which are meteorites whose parent bodies are thought to have formed in the inner solar system. 
The carbon content of these types of chondrites is $\sim 0.1~\rm wt\%$ \citep{Jarosewich1990Chemical-analys}. This value is much lower than that of IDPs, and would be more consistent with that of the bulk Earth if we hypothesize that the Earth's core (which constitutes 1/3 of the Earth mass) contains $1~\rm wt\%$ carbon \citep{Wood2013Carbon-in-the-C}.
Moreover, \citet{Gail2017Spatial-distrib} have recently suggested that the materials comprising these chondrites could have formed from initially carbon-rich dust. They show that the flash heating events that produced chondrules, mm-sized spherules that dominate the ordinary and enstatite chondrites, were able to destroy most carbonaceous materials contained in the precursor dust aggregates. 
This does not mean that all OMG aggregates would have lost organic mantles before they formed planetesimals, because the the duration of chondrule formation ($\approx 3~\rm Myr$; \citealt{Connelly2012The-Absolute-Ch}) is much longer than the timescale of dust growth into planetesimals ($\approx 10^3~\rm yr$; see the left panel of Figure~\ref{f:grow32}). 

\begin{figure*}[t]
\centering
 \resizebox{18cm}{!}{\includegraphics{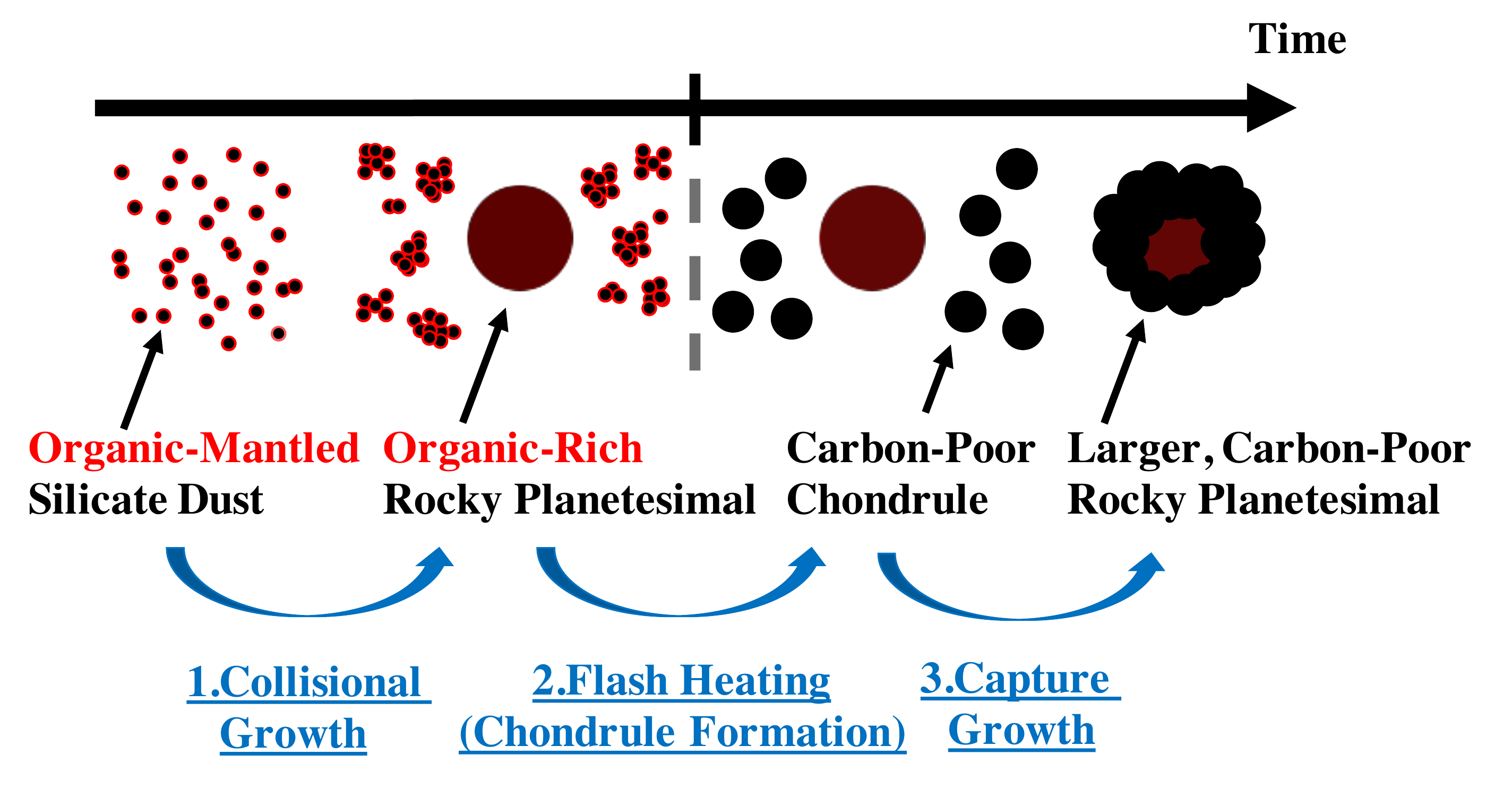}}
\caption{Possible pathway for rocky planet formation aided by organics. See Section \ref{ss:imp} for explanation of the three steps.}
\label{f:scenario}
\end{figure*}
Therefore, we can speculate a scenario in which the ``seeds'' of planetesimals form promptly with the aid of organics and then grow into carbon-poor planetesimals and protoplanets through carbon-poor chondrule accretion.
We postulate that the dust grains in the inner region of the solar nebula had organic mantles similar to those of CP IDPs before the grains formed large aggregates. We then envisage that the OMGs grew into carbon-poor solid bodies in the following three steps (Figure \ref{f:scenario}):

\begin{enumerate}

    \item The OMGs in a disk annulus of $T\sim 200$--400 K quickly grew into planetesimals thanks to the soft, sticky organic mantles. Outside this annulus, dust grains did form macroscopic aggregates up to $\sim 1$ m in size, but collisional fragmentation inhibited their further growth into planetesimals because their organic mantles were either not soft enough (for $T \la 200~\rm K$) or absent (for $T \ga 400~\rm K$).
  
   \item Repeated heating events converted the organic-rich aggregates into carbon-poor chondrules. A substantial fraction of the organic matter, including organics, inside the aggregates was lost during these events. The chondrules that formed from the macroscopic aggregates outside the seed planetesimal annulus were then gradually implanted into the annulus through radial inward drift and perhaps turbulent diffusion.
    
    \item  The seed planetesimals accreted the implanted carbon-poor chondrules, forming rocky planetesimals that were larger but less carbon-rich than the seed planetesimals. Most of the carbon-poor planetesimals grew into the terrestrial planets and/or their embryos (protoplanets) through planetesimal--planetesimal collisions and/or further accretion of chondrules \citep{Johansen2015Growth-of-aster,Levison2015Growing-the-ter}.
    The remnants might be the parent bodies of the ordinary and enstatite chondrites. 
    
\end{enumerate}

 In this scenario, the final carbon content of rocky planetesimals would depend on whether they only capture chondrules or capture chondrites, mixtures of chondrules and fine-grained matrices, because matrix grains are the dominant carbon carriers in chondrites \citep[e.g.,][]{Makjanic1993Carbon-in-the-m,Alexander2007The-origin-and-}. In the latter case, the final carbon abundance of large planetesimals could be as high as the bulk carbon content of chondrites, $\sim$ 0.1--1 wt${\%}$ \citep{Jarosewich1990Chemical-analys}.
 The final carbon content may also have a spatial variation depending on how deep inside the planetesimal-forming zone chondrules can be implanted. However, the spatial variation would be small if the planetesimal belt that formed the solar-system terrestrial planets were narrow as suggested by  \citet{Kokubo2006Formation-of-Te} and \citet{Hansen2009Formation-of-th}.

Our scenario requires heating events that can efficiently convert organic-rich aggregates outside the planetesimal-forming zone into chondrules. The total mass of the  chondrules must have been higher than that of initially carbon-rich planetesimals, and than that of the carbon-rich aggregates that survived the heating events. 
The latter condition is necessary because planetesimals can in principle accrete both chondrules and carbon-rich aggregates. 
The question is what mechanism was responsible for such heating events. Possible mechanisms for chondrule-forming heating events proposed so far include bow shocks around eccentric planetesimals/protoplanets \citep{Morris2018Formation-of-Ch}, bow shocks associated with spiral density waves in the solar nebula \citep{Desch2002A-model-of-the-}, heating by lightning discharge \citep{Horanyi1995Chondrule-forma,Johansen2018Harvesting-the-}, radiative heating by hot planetesimals \citep{Herbst2016A-new-mechanism,Herbst2019A-Radiative-Hea}, and planetesimal collisions \citep{Johnson2015Impact-jetting-}.
Our scenario prefers mechanisms that can form millimeter-sized chondrules from 
meter-sized aggregates, which dominate the total mass of dust outside the planetesimal-forming zone in our simulation.
Heating by shock waves is one mechanism that could produce millimeter-sized melt droplets from larger aggregates \citep{Susa2002On-the-Maximal-,Kadono2005Breakup-of-liqu,Kato2006Maximal-size-of}. The radiative heating model by \citet{Herbst2019A-Radiative-Hea} also invokes meter-sized aggregates as the source of chondrules, but the amount of the produced chondrules may not be high enough to meet our requirement. We cannot rule out that other mechanisms are also compatible with our scenario.  

The scenario proposed above is speculative and needs to be tested in future work. First of all, there is no observational support for the assumption that thick organic mantles were present on the grains in the inner region of the solar nebula. The scenario also assumes that there was a continuous supply of chondrules, and hence their precursor dust aggregates, so that the final planetesimals were dominantly composed of the carbon-poor chondrules. 
Assessment of these assumptions will require modeling of the global transport of chondrules and precursor dust aggregates combined with realistic models for organic synthesis and chondrule formation in the solar nebula.

\section{Summary} \label{s:summary}
We have explored the possibility that rocky dust grains with organic mantles grow into rocky planetesimals through mutual collisions inside the water snow line of protoplanetary disks.
We constructed a simple adhesion model for organic-mantled grains by modifying a contact model for uniform elastic spheres 
(Section \ref{s:adhesion}).
In general, the stickiness of uniform elastic grains with surface adhesion forces is high when the grains are soft, simply because softer particles make a larger contact area \citep{Johnson301}.
Organic matter is sticky in this respect because they have low elasticity, in particular in warm environments \citep[][see also our Figure \ref{f:Gor}]{Kudo2002The-role-of-sti}.  
In the case of OMGs, however, the hard silicate core limits the size of the contact area, and therefore the thickness of the organic mantle relative to the grain radius, $\Delta R_{\rm or}/R$, enters as an important parameter that determines the maximum stickiness of OMGs (Figure~\ref{f:2l}).
Our adhesion model shows that aggregates of $0.1 ~\micron$-sized OMGs still can overcome the fragmentation barrier in weakly turbulent protoplanetary disks if 
$\Delta R_{\rm or}/R \ga 0.03$ and if the temperature is above $\approx 220\rn K$ (Figures \ref{f:vfrag} and \ref{f:monomer1}).

Using the adhesion model, we also simulated the global collisional evolution of OMG aggregates inside the snow line of a protoplanetary disk (Section \ref{s:simulation}).
Our simulations demonstrate that OMG aggregates with $R = 0.1~\micron$ and $\Delta R_{\rm or}/R = 0.03$ can overcome the fragmentation barrier and form rocky planetesimals at $r \la 1 \rn au$ (the left panel of Figure \ref{f:grow32}). At these orbits, the radial drift barrier is also overcome thanks to the aerodynamic properties of the drifting aggregates in this high-density region \citep{Birnstiel2010Gas--and-dust-e,Okuzumi2012Rapid-Coagulati,Drc-azkowska2014Rapid-planetesi}. Because organic matter is likely unstable at $T \ga 400~\rm K$ \citep{Kouchi2002Rapid-Growth-of}, one can only expect planetesimal formation by the direct coagulation of OMGs in a temperature range $200~{\rm K} \la T \la 400~\rm K$. Such   a   narrow   planetesimal-forming zone can provide favorable conditions for the formation of  the  terrestrial  planets  in  the  solar  system (Section~\ref{ss:imp}).

It is not obvious if our findings can explain the formation of the solar-system terrestrial planets, because their carbon content is low (Section \ref{ss:how}).
The organic-rich rocky planetesimals produced in our simulations have a high carbon content of $\sim$ 5 wt\%, which is too high to be consistent with the carbon content of the present-day Earth, even if the Earth's core contains 1 wt\% carbon.
Thus, we proposed a scenario in which sticky OMGs form the seeds of planetesimals and then grow into larger, less carbon-rich planetesimals through the accretion of carbon-poor (ordinary and enstatite) chondrules (Figure \ref{f:scenario}). 
These carbon-poor chondrules could also form from OMG aggregates because most carbonaceous materials contained in the precursor aggregates can be destroyed during flash heating events \citep{Gail2017Spatial-distrib}.
The proposed scenario for terrestrial planet formation is still highly speculative and further assessments are needed in future work. 

\acknowledgments 
We thank Hidekazu Tanaka, Akira Kouchi, Hiroko Nagahara, Joanna Dr\c{a}\.{z}kowska, Shoji Mori, Kazumasa Ohno, Haruka Sakuraba, and Sota Arakawa for useful comments and discussions. This work was supported by JSPS KAKENHI Grant Numbers JP16H04081, JP16K17661, JP18H05438, and JP19K03926.


\end{document}